\documentclass{elsart}

\usepackage{amssymb}
\usepackage{amsmath}

\newcommand{\bra}{\left\langle}
\newcommand{\ket}{\right\rangle}

\newcommand{\ketsE}{\right\rangle^{{\rm st}}_{E}}
\newcommand{\ketsEz}{\right\rangle^{{\rm st}}_{E=0}}
\newcommand{\kets}{\right\rangle^{{\rm st}}}

\newcommand{\kettE}{\right\rangle^{{\rm tr}}_{E}}
\newcommand{\kettV}{\right\rangle^{{\rm tr}}_{V}}

\newcommand{\ps}{p^{\rm st}}
\newcommand{\pc}{p^{\rm can}}

\newcommand{\Ps}{P^{{\rm st}}}
\newcommand{\PtE}{P^{\rm tr}_{E}}

\newcommand{\Pgc}{P^{\rm tr}_{\Delta \mu}} 
\newcommand{\Pgct}{P^{{\rm tr}*}_{\Delta \mu}}
\newcommand{\pgc}{p^{\rm gcan}}

\newcommand{\pder}[2]{\frac{\partial #1}{\partial  #2}}

\newcommand{\der}[2]{\frac{d #1}{d  #2}}

\newcommand{\Dc}{D_{\rm c}}
\newcommand{\mul}{\mu_{\rm 1}}
\newcommand{\mur}{\mu_{\rm L}}
\newcommand{\cl}{c_{\rm 1}}
\newcommand{\crr}{c_{\rm L}}
\newcommand{\cb}{c_{\rm bulk}}
\newcommand{\Tl}{T_{\rm 1}}
\newcommand{\Tr}{T_{\rm L}}
\newcommand{\Tb}{T_{\rm bulk}}
\newcommand{\Phil}{\Phi_{\rm 1}}
\newcommand{\Phir}{\Phi_{\rm L}}

\newcommand{\vecI}{{\boldsymbol I}}
\newcommand{\vecalpha}{{\boldsymbol \alpha}}
\newcommand{\vecr}{{\boldsymbol r}}
\newcommand{\vece}{{\boldsymbol e}}
\newcommand{\vecxi}{{\boldsymbol \xi}}
\newcommand{\vecF}{{\boldsymbol F}}
\newcommand{\dt}{\Delta t}
\newcommand{\vecGamma}{{\boldsymbol \Gamma}}

\newcommand{\veceta}{{\boldsymbol \eta}}

\begin{document}

\begin{frontmatter}



\title{Linear response theory in stochastic many-body systems 
revisited}  
  

\author{Kumiko Hayashi and Shin-ichi Sasa}

\address
{Department of Pure and Applied Sciences,  
University of Tokyo, Komaba, Tokyo 153-8902, Japan}

\begin{abstract}
The Green-Kubo relation, the Einstein relation, and the 
fluctuation-response relation are representative universal 
relations among measurable quantities that are valid in the 
linear response regime.  We provide  pedagogical proofs of these  
universal relations for stochastic many-body systems.  Through 
these simple proofs, we characterize  the three relations as follows.  
The Green-Kubo relation is a direct result of the local detailed 
balance condition, the fluctuation-response relation represents 
the dynamic extension of both the Green-Kubo relation  and the 
fluctuation relation in equilibrium statistical mechanics, and the 
Einstein relation can be understood by considering thermodynamics.
We also clarify the interrelationships among the universal relations.
\end{abstract}

\end{frontmatter}


\section{Introduction}\label{s:model:intro}


Let us consider the general class of systems exhibiting the 
macroscopic flow of some quantity. Familiar examples include 
electric  conduction systems, heat conduction systems and driven 
colloidal systems.  Such systems often exist in  
states in which the current is stationary.  We refer to these 
states as {\it nonequilibrium  steady states}.  Nonequilibrium 
steady  states are the simplest type of  nonequilibrium states.  
However, even  for the class consisting of such simple states,  
there presently exists  no universal framework to calculate 
macroscopic properties of  systems from microscopic information, 
in contrast to the class of equilibrium states,  for  which 
macroscopic properties can be  calculated from the partition function of 
the system, employing the  well-established theory of 
statistical mechanics. 


At the present time, there exists a theory of 
nonequilibrium systems only for the regime very near equilibrium. 
This so-called {\it linear response theory} consists of a  set 
of  universal relations, which includes the Green-Kubo relation,   
the fluctuation-response relation and  the Einstein-relation.  
These relations relate the dynamical properties of fluctuations 
in systems under equilibrium conditions to linear transport  
properties of nonequilibrium  systems. The validity of these 
relations has been confirmed  by both experimental and theoretical 
investigations \cite{kubohasi}.


In recent years, properties regarding fluctuations and  linear 
responses  to perturbations have been investigated for 
nonequilibrium states  even outside the linear response regime, 
in steady state systems \cite{g.eyink}, aging systems 
\cite{g1,g2,g3} and other systems \cite{Ono,Kolton}.   Obviously, 
we cannot expect the relations proved within  linear response 
theory to be  generally valid outside the linear response regime. 
However, there have been several relations proposed and investigated 
recently that represent extensions of the relations in  linear 
response theory to systems far from equilibrium 
\cite{HSIII,HSIV,HSV,KH,HHS,Harada,HaradaSasa}.

We wish to acquire a systematic understanding of these extended 
relations among fluctuations and response properties, aiming at 
the construction of  a statistical mechanics of  nonequilibrium 
systems far from equilibrium.  For this purpose, first it is 
necessary to grasp the essence of the basic framework of linear 
response theory, in  particular focusing on the elucidation of 
the interrelations among  the linear response relations, 
thermodynamics and equilibrium statistical mechanics.


Of course, all of the relations in linear response theory 
can be derived  by applying a formal perturbative expansion to 
quantum or classical mechanical descriptions on the basis 
of equilibrium statistical mechanics 
\cite{Green,CallenWelton,KuboTomita,Nakano,Kubo}.  However, such 
formal derivations of these relations does not illuminate the 
physical picture (see Section \ref{s:dis}).  
In this paper, instead of applying a formal perturbation theory to 
quantum or  classical mechanical systems,   we attempt to elucidate 
the essence of linear response theory by studying stochastic 
many-body systems.


Among the various types of stochastic many-body systems, we choose 
to investigate two kinds of one-dimensional nonequilibrium lattice 
gases in which particles hop stochastically  from one lattice site 
to another according to a given transition rate \cite{Spohn,KLS,LG1,LG2}. 
Specifically,  we study  a driven lattice gas, in which particles 
are driven by an  external  driving force in the bulk, and a 
boundary-driven lattice gas, in which particles are driven by the 
difference in  chemical potential between two particle reservoirs 
in contact with  the boundaries.  Although we choose these 
models for their mathematical simplicity, it is straightforward to 
apply the method of proof to other stochastic many-body systems, 
such as Langevin systems that describe the motion of many beads 
in a liquid. 


This paper is organized as follows. In Section \ref{s:assmp}, 
we state our physical point of view and the basic assumptions used 
in the derivation  of the linear response relations.  In Section 
\ref{s:model}, we introduce  a one-dimensional driven lattice gas.  
In Section \ref{s:gk}, we present a simple derivation of the 
Green-Kubo relation, employing  a key identity, with which various 
universal relations characterizing nonequilibrium steady states can be 
derived. In Section \ref{s:frr}, we derive the fluctuation-response 
relations and elucidate its connection to  equilibrium statistical 
mechanics.  In Section \ref{s:ein}, we derive the Einstein relation 
for the model, and  clarify the connections among these three 
universal relations and thermodynamics. In Section \ref{s:b}, we 
derive the Green-Kubo relation for a boundary-driven lattice gas, 
which is a different type of nonequilibrium lattice gas,  in order 
to demonstrate  the generality  of our argument. Finally, in 
Section \ref{s:dis}, we give several  important remarks concerning 
linear response theory.

\section{Basic assumption}\label{s:assmp}

Let us recall that the macroscopic properties of an equilibrium  
system  can be investigated with statistical mechanics 
using an effective  multi-dimensional variable $\vecalpha$ and 
the corresponding effective Hamiltonian $H(\vecalpha)$.  
More precisely,  in order to obtain such a description,    
the variable $\vecalpha$, described  statistical mechanically,   
need not constitute a complete set of mechanical variables 
that specifies the  microscopic state in classical or  quantum 
systems.  For  example, consider a system consisting  of many beads 
suspended in a liquid. In this case, $\vecalpha$ can be regarded as 
representing the set of  centers of mass of the beads, with 
$H(\vecalpha)$ describing the effective interactions between 
them. The effect of the liquid  can be included in this effective 
Hamiltonian $H(\vecalpha)$. 

Here, we suppose that the set of variables $(\vecalpha, \vecGamma)$ 
provides a complete mechanical description of the system under 
consideration. For the bead system, $\vecGamma$ is defined so 
that $(\vecalpha, \vecGamma)$ represents all the positions and  
conjugate momenta of the atoms composing the beads and the liquid  
in an adiabatic container. 

Now, we consider a nonequilibrium system which is such that  the 
degrees of freedom represented by  $\vecGamma$ remain in equilibrium, 
while only those represented by  $\vecalpha$ are in  a nonequilibrium 
state. It is not certain that such an assumption would be valid, 
and in general the degree of error it introduces would depend on 
the system we consider. Physically, this assumption seems reasonable 
in the case that  the typical time scale of $\vecalpha$, $\tau_\alpha$,  
is much longer than that of $\vecGamma$, $\tau_\Gamma$.  We consider 
such a case and introduce a time interval $\Delta t$ satisfying  
$\tau_\Gamma \ll \Delta t \ll \tau_\alpha$.  In the argument below, 
we refer  the subsystem composed of  the degrees of freedom $\vecGamma$  
as the {\it heat bath}.  Because this heat bath is assumed to be always  
in equilibrium, its temperature, $T$, is defined for nonequilibrium 
states of the remaining degree of freedom.  

In order to describe the nonequilibrium states of such systems,  
we need to determine the time evolution of the variable $\vecalpha$ 
on the time scale of the interval $\Delta t$.   Due to the separation 
of  time scales represented  by the relation $\tau_\Gamma \ll \Delta t$,  
it is reasonable to  assume that  $\vecalpha$ is described by a 
Markov stochastic process with a transition probability  
$T(\vecalpha \to \vecalpha')$.

The fundamental assumption in this paper is expressed by 
the local detailed balance 
condition
\begin{equation}
\frac{T(\vecalpha \to \vecalpha')}{T(\vecalpha' \to \vecalpha)}
=\e^{-\beta Q(\vecalpha \to \vecalpha')}
\label{LDB:alpha}
\end{equation}
where $\beta=1/T$.  Here, the Boltzmann constant is set to unity. 
The quantity  $Q(\vecalpha \to \vecalpha')$  represents the energy 
transferred from this heat bath, which we call heat, during  the 
time evolution symbolized by $\vecalpha \to \vecalpha'$ taking place 
over the time interval $\Delta t$. 

In the system under equilibrium conditions, because the energy 
conservation expressed as 
 $Q(\vecalpha \to \vecalpha')=H(\vecalpha')-H(\vecalpha)$ 
should hold, (\ref{LDB:alpha}) becomes the detailed balance condition 
of the transition probability:
\begin{equation}
\pc(\vecalpha)T(\vecalpha \to \vecalpha')
=\pc(\vecalpha')T(\vecalpha' \to \vecalpha),
\label{DB:alpha}
\end{equation}
with respect to the canonical distribution
\begin{equation}
\pc(\vecalpha)=\frac{e^{-\beta H(\vecalpha)}}{Z}, 
\label{pcan:alpha}
\end{equation}
where $Z$ is a normalization constant. Combining (\ref{DB:alpha}) 
with the normalization condition of the probability,   
$\sum_{\vecalpha'} T(\vecalpha \to \vecalpha')=1$, we derive 
\begin{equation}
p^{\rm can}(\vecalpha)=
\sum_{\vecalpha'} p^{\rm can}(\vecalpha')T(\vecalpha' \to \vecalpha).
\label{ss:alpha}
\end{equation}
This equality implies that (\ref{pcan:alpha}) is the stationary 
distribution of $\vecalpha$. This result is consistent with 
equilibrium statistical mechanics. Also, it turns out that the 
detailed balance condition (\ref{DB:alpha}) implies  time-reversal 
symmetry in the sense that the two-time joint probability 
$p^{\rm can}(\vecalpha)T(\vecalpha \to \vecalpha')$ is invariant
under the exchange of $\vecalpha$ and $\vecalpha'$. 

By contrast, in the case of a nonequilibrium state, it is not easy to 
obtain  a physical interpretation of (\ref{LDB:alpha}). We explain the 
origin of the name {\it local detailed balance condition}  in 
Section \ref{s:model},   and  demonstrate that all the universal 
relations can be derived from this assumption in Sections 
\ref{s:gk}, \ref{s:frr}, and \ref{s:ein}.  In  the remainder of 
this section,   we explain how the local detailed balance 
condition  (\ref{LDB:alpha})  places restrictions on the allowed 
types of interactions  with the heat bath.

In order to demonstrate this point explicitly, we consider a 
many-body bead system in which an external driving force is applied 
to the beads.  We wish to derive the transition probability 
$T(\{\vecr_j\} \to \{\vecr_j'\})$  over an infinitesimally small 
time interval $\Delta t$, using the local detailed balance condition, 
where  $\vecr_i$ ($i=1,\cdots, N$) represents the position of the  
$i$-th bead.  

First, note that the mechanical  force acting on the $i$-th bead is 
given by   
\begin{equation}
\vecF_i(\{\vecr_j\}) =
f\vece_x-\pder{U_0(\vecr_i)}{\vecr_i}
-\sum_{j=1;j\not =i }^N \pder{U(\vecr_i-\vecr_j)}{\vecr_i},
\end{equation} 
where $U_0$ is a one-body potential (e.g. a spatially periodic 
potential), $U$ is the interaction potential between beads, and 
$f$ is an external driving force.  Then, because the change in  
momentum is much smaller than the dissipative force in this situation, 
the averaged motion caused by the force  $\vecF_i(\{\vecr_j\})$ 
is determined by the force balance equation 
\begin{equation}
-\gamma (\bra \vecr_i'\ket -\bra \vecr_i\ket)+\bar \vecF_i \dt=0,
\end{equation}
where $\gamma$ represents the dissipation constant,  and 
$\bar \vecF_i$ is the average of $\vecF_i$  during the 
transition, given by 
\begin{equation}
\bar \vecF_i =
\frac{1}{2}[\vecF_i(\{\vecr_j(t)\})+\vecF_i(\{\vecr_j(t+\dt)\})].
\end{equation} 

Under the influence of the interactions with  molecules in the 
liquid,   the positions of  the  beads  fluctuate  around this 
averaged motion. Because we choose  $\Delta t$  to be much larger 
than the time scale characterizing the dynamics of  the molecules,  
we expect that these fluctuations exhibit a  Gaussian distribution 
with a  dispersion proportional to $\Delta t^{-1}$. That is, it 
is reasonable to express the fluctuations by the transition probability 
\begin{equation}
T(\{\vecr_j\} \to \{\vecr_j'\})
= \frac{1}{Y}
\e^{-\Delta t \frac{1}{4b}\sum_{i=1}^N
|\gamma \dot \vecr_i -\bar \vecF_i|^2 },
\label{trans:bead}
\end{equation}
where $Y$ is a normalization factor, $b$ is a parameter which 
represents the strength of the interaction with the heat bath, and 
$\dot \vecr_i$  should be interpreted  as
\begin{equation}
\dot \vecr_i=\frac{\vecr_i'-\vecr_i}{\Delta t}.
\end{equation}
We wish to determine $b$ in (\ref{trans:bead}) from the local 
detailed balance condition. In order to do this, we consider the 
following ratio 
of  transition probabilities: 
\begin{equation}
\frac{T(\{\vecr_j\} \to \{\vecr_j'\})}{T(\{\vecr_j'\} \to \{\vecr_j\})}
= 
\e^{\Delta t \frac{\gamma}{b}  \sum_{i=1}^N \dot \vecr_i \bar \vecF_i}.
\label{LDP:bead}
\end{equation}
In this expression, it should be noted that  the quantity 
$\Delta t \sum_{i=1}^N\dot \vecr_i \bar \vecF_i$ is the work done 
by the force $\bar \vecF_i$, which is  equal to the heat dissipated 
into the heat bath \cite{sekimoto}  during the transition from  
$\{\vecr_i\} $  to $\{\vecr_i'\}$.  Then, by comparing (\ref{LDP:bead}) 
with the local detailed balance condition (\ref{LDB:alpha}),  
we find the relation  
\begin{equation}
b=\gamma T. 
\label{2nd}
\end{equation}
In this way, we have determined the transition probability  
(\ref{trans:bead}), with  the explicit form of $b$ from the local 
detailed balance condition (\ref{LDB:alpha}). 

Note that the stochastic process given by the transition probability 
(\ref{trans:bead}) can be expressed as  
\begin{equation}
\gamma (\vecr_i'-\vecr_i) =\bar \vecF_i \dt+\Delta W_i, 
\label{model:brown}
\end{equation}
where $\Delta W_i$ represents Brownian motion satisfying
\begin{equation}
\bra \Delta W_i \Delta W_j \ket = 2 b \delta_{ij}\Delta t.
\end{equation}
Then, taking the limit $\dt \to 0$, we obtain the Langevin  equation
\begin{equation}
\gamma \dot \vecr_i =f\vece_x- \pder{U_0(\vecr_i)}{\vecr_i}
-\sum_{j=1;j\not =i }^N \pder{U(\vecr_i-\vecr_j)}{\vecr_i}+\vecxi_i ,
\label{model:bead}
\end{equation}
where $\vecxi_i$ represents zero-mean Gaussian white noise with 
\begin{equation}
\bra \vecxi_{i}(t)\vecxi_{j}(t') \ket=
2 b \delta_{ij}\vecI \delta(t-t').
\label{noise}
\end{equation}
Here, $\vecI$ is a unit matrix.   The condition (\ref{2nd}) is called  
the {\it fluctuation-dissipation relation of the second kind} for 
the Langevin equation (\ref{model:bead}) with (\ref{noise}).

\section{Model}\label{s:model}

We can derive all the universal relations in linear response
theory for the Langevin equation (\ref{model:bead}) with (\ref{2nd}) 
and (\ref{noise}).  However, this derivation is complicated when 
mathematical rigor is maintained. Instead of the Langevin model, we 
study a lattice gas,  which also serves as a model of many-body 
beads systems, to avoid the complicated analysis that is irrelevant 
to the essence of  linear response theory.  

In a lattice gas,  particles   hop stochastically  from one lattice 
site to another at a given transition rate. Particularly, when  
two or more particles cannot be placed  at one site, the equilibrium 
state is described by a Hamiltonian  which is  equivalent to that 
for a spin model with a variable transformation. For mathematical 
simplicity, the form of the Hamiltonian usually corresponds 
to the Ising model.  In Subsection \ref{s:model:LG}, we explain the 
one-dimensional equilibrium lattice gas.  A two or higher 
dimensional model can be easily interpreted from the one-dimensional 
case. 

A {\it driven} lattice gas,  introduced  in Subsection 
\ref{s:model:DLG},  is defined  by minimally  modifying  to the 
equilibrium lattice gas  so that it exhibits nonequilibrium steady 
states.  The model was numerically and theoretically  investigated 
in Refs. \cite{KLS,LG1,LG2}.

\subsection{One-dimensional equilibrium  lattice gas}\label{s:model:LG}

We first define an occupation variable, $\eta_x$, on each site $x$ 
in a one-dimensional periodic lattice $x=1,2, \cdots, L$.  $\eta_x=1$ 
when the site $x$ is occupied by a particle  and $\eta_x=0$  if 
$x$ is unoccupied.  A  periodic boundary condition is imposed by 
introducing a boundary site at $x=0$ and setting $\eta_0=\eta_L$.    
The collection of all occupation variables $(\eta_1, \cdots, \eta_L)$ 
represents the particle positions in the system.  We denote  
$(\eta_1, \cdots, \eta_L)$ as $\veceta$, which is referred to as 
 the {\it configuration}.  For convenience, we use $\veceta^x$  
to represent the configuration when the value of $\eta_x$ is replaced 
with that of $1-\eta_x$.  In a similar way, $\veceta^{xy}$ 
represents the configuration when the values of $\eta_x$ 
and $\eta_y$ are replaced by $1-\eta_x$ and $1-\eta_y$. 
(Note that the configuration after exchanging the value of $\eta_x$ 
and $\eta_y$ is represented by this $\veceta^{xy}$.)  
The interaction  between the particles is described by the Hamiltonian 
\begin{equation}
H_0( \veceta )= - \sum_{x=0}^{L-1} \eta_x \eta_{x+1}.
\label{eqhami}
\end{equation}

The time evolution of $\veceta$ is expressed as follows: 
At each time step, a nearest neighbor pair $x$ and $y$, 
$0 \le x,y \le L $, is randomly selected and  the values of 
$\eta_x$ and $\eta_y$  are exchanged using  the exchange probability  
$c_0(x,y;\veceta)$  that satisfies 
\begin{equation}
c_0(x,y;\veceta)=c_0(x,y;\veceta^{xy})
\e^{-\beta  Q_0(\veceta\to\veceta^{xy})},  
\label{ex0}
\end{equation}
with 
\begin{equation}
Q_0( \veceta \to \veceta')= H_0( \veceta' )-H_0(\veceta). 
\label{q0}
\end{equation}
$Q_0(\veceta \to \veceta^{xy})$ corresponds to the heat absorbed  
from the heat bath  for the configuration change 
$\veceta \to \veceta^{xy}$.  We also define $c_0(x,y;\veceta)=0$ 
for pairs $x $ and $y$  such that $|x-y|\not =1$.  We regard this 
step as a Monte Carlo step.

Examples of common concrete forms of  $c_0(x,y;\veceta)$  are the 
Metropolis method 
\begin{equation}
c_0(x,y;\veceta)={\rm min}\{1, \e^{-\beta Q_0(\veceta\to\veceta^{xy})}\},
\label{mt}
\end{equation}
the heat bath method 
\begin{equation}
c_0(x,y;\veceta)=\frac{1}{1+ \e^{\beta Q_0(\veceta\to\veceta^{xy})}},
\label{hb}
\end{equation}
and the exponential rule
\begin{equation}
c_0(x,y;\veceta)= {\rm  const.}
\e^{-\frac{1}{2}\beta Q_0(\veceta\to\veceta^{xy})}, 
\label{sym}
\end{equation}
where  const. is chosen so as to satisfy the condition that 
$c_0(x,y;\veceta)\le 1$.  It can be easily checked that (\ref{mt}),  
(\ref{hb}) and (\ref{sym}) satisfy (\ref{ex0}).  

The transition probability $T_0(\veceta \to \veceta')$ for one 
Monte Carlo step is written as
\begin{equation}
T_0(\veceta \to \veceta') = \frac{1}{L} c_0(x,y;\veceta)
\label{tran0}
\end{equation}
for $\veceta'=\veceta^{xy}\not =\veceta$, and $T_0(\veceta \to \veceta)$ 
is determined by the normalization condition of the probability 
\begin{equation}
\sum_{\veceta'}T_0(\veceta \to \veceta')=1.
\label{unity}
\end{equation}
Note that  in the argument, $T_0(\veceta \to \veceta')$ is 
different from  $c_0(x,y;\veceta)$; the transition probability,  
$T_0(\veceta \to \veceta')$, is a function of two configurations, while 
the exchange probability, $c_0(x,y;\veceta)$, is a function of the 
sites $x,y$ and the configuration $\veceta$. 

{}From (\ref{ex0}) and (\ref{tran0}), we find that $T_0(\veceta \to 
\veceta')$ satisfies the local detailed balance condition
\begin{equation}
\frac{T_0(\veceta \to \veceta')}{T_0(\veceta' \to \veceta)}
=\e^{-\beta Q_0(\veceta \to \veceta')}.
\label{LDB0}
\end{equation}  
Here, using (\ref{q0}) in this equilibrium case,  the condition 
(\ref{LDB0}) becomes the detailed balance condition 
\begin{equation}
\pc(\veceta)T_0(\veceta \to \veceta')
=\pc(\veceta')T_0(\veceta' \to \veceta), 
\label{DB}
\end{equation}
with respect to the canonical distribution
\begin{equation}
\pc(\veceta)=\frac{e^{-\beta H_0(\veceta)}}{Z}\delta(\sum_{x=1}^L\eta_x-N). 
\label{pcan}
\end{equation}
Here, $Z$ is a normalization constant, and $N$ is the total number of  
particles,
\begin{equation}
N=\sum_{x=1}^L\eta_x,
\label{num0}
\end{equation} 
which is conserved through this time evolution.   
$\bar \rho=N/L$ is introduced to be a parameter of the model. 
Similar to Section \ref{s:assmp}, using (\ref{DB}),   
we can confirm  that (\ref{pcan}) is the stationary distribution 
of this model in the equilibrium case.

\subsection{One-dimensional driven lattice gas}\label{s:model:DLG}

In this subsection, a {\it driven} lattice gas that 
describes a physical situation where the particles are driven 
by a uniform driving force $E$ in the bulk is introduced 
as a simple extended model of the equilibrium lattice gas.   
First,  we  define the  net number of particles that  
hop from $x$ to $x+1$ in the configuration change 
$\veceta \to \veceta'$ as 
\begin{equation}
\Phi_x(\veceta \to \veceta')=
\eta_x(1-\eta'_{x})\eta'_{x+1}(1-\eta_{x+1})-
\eta_x'(1-\eta_x)\eta_{x+1}(1-\eta'_{x+1}).
\label{phix}
\end{equation}
Note that $\Phi_x(\veceta \to \veceta')$ takes a value in  
$\{\pm 1, 0\}$. Then, the heat $Q_0$ given in (\ref{q0}) is 
replaced with  
\begin{equation}
Q( \veceta \to \veceta')
= H_0( \veceta' )-H_0(\veceta) 
- E \sum_{x=0}^{L-1} \Phi_x(\veceta \to \veceta'),
\label{derh}
\end{equation}
to consider the effect of the driving force on the time evolution 
rule, because  $Q( \veceta \to \veceta')$ is interpreted as
the heat  absorbed from the heat bath  for the configuration 
change $\veceta \to \veceta'$ in one Monte Carlo step. 
Using this replacement, $c_0(x,y;\veceta)$ in (\ref{ex0}) and 
$T_0(\veceta \to \veceta') $ in (\ref{tran0}) are replaced by  
$c(x,y;\veceta)$ and $T(\veceta \to \veceta')$, respectively. 
Then, $T(\veceta \to \veceta')$ satisfies  the local detailed 
balance condition
\begin{equation}
\frac{T(\veceta \to \veceta')}{T(\veceta' \to \veceta)}
=\e^{-\beta Q(\veceta \to \veceta')}.
\label{LDB}
\end{equation}  

Here, one may consider whether  $Q(\veceta \to \veceta')$ given in 
(\ref{derh}) can be expressed as 
\begin{equation}
Q(\veceta \to \veceta')=\sum_{x=0}^{L-1}  h_x(\veceta')-
\sum_{x=0}^{L-1}  h_x(\veceta),
\label{Qh}
\end{equation}
with a local Hamiltonian 
\begin{equation}
h_x ( \veceta )=   -  \eta_x \eta_{x+1}-E  {x \eta_x}. 
\label{localhami}  
\end{equation}
If  (\ref{Qh}) and (\ref{localhami}) were valid, 
this should correspond to the detailed balance condition, because 
(\ref{LDB}) can be written as
\begin{equation}
 T(\veceta \to \veceta')\e^{-\beta \sum_{x=0}^{L-1} h_x(\veceta)}
=T(\veceta' \to \veceta) \e^{-\beta \sum_{x=0}^{L-1}  h_x(\veceta')}.
\label{LDB2}
\end{equation}
However, (\ref{Qh}) is valid only when  $\veceta'=
\veceta^{xx+1}$ with $ 0 \le x \le L-2$, and is invalid when  
$x=L-1$, unless $E=0$. 
More generally, when  $E\not=0$, 
(\ref{Qh}) can be satisfied for the configuration 
changes with one exception, but not for all exchanges,  even if 
we introduce another local Hamiltonian.   Due to this property,  
the condition (\ref{LDB}) is called the  {\it local detailed 
balance condition}. {}As shown in the  above argument, the local 
detailed balance condition   does not necessarily imply the 
detailed balance condition,  except for the equilibrium case.

Throughout this paper, we regard  $L$ Monte Carlo steps as the 
unit of time, which is the number of Monte Carlo  steps  
necessary for all sites to be  selected once on the average. 
This unit of time is called one {\it MCS} (Monte Carlo step per site). 
Then, we denote the history of the configurations from $0$ to 
$\tau$ MCS as $[\veceta]=(\veceta(0), \veceta(1),\cdots, 
\veceta(L\tau))$. The probability of a history $[\veceta]$ in 
the nonequilibrium steady state, $\Ps_E([\veceta])$, is written as  
\begin{equation}
\Ps_E([\veceta])=\ps_{E}(\veceta(0))
T(\veceta(0)\to\veceta(1))\cdots T(\veceta(L\tau-1) \to \veceta(L\tau)),
\end{equation}
where $\ps_{E}(\veceta)$ is the stationary distribution in  the 
system with the driving force $E$.  Specifically, when  $E=0$, 
$\ps_{E=0,N}(\veceta)$ is identical to $\pc(\veceta)$ defined 
in (\ref{pcan}).  Then, the statistical average of a history 
dependent quantity $A$ in the steady state is denoted  as
\begin{equation}
\bra A\ketsE\equiv\sum_{[\veceta]}\Ps_E([\veceta]) A([\veceta]).
\label{aa}
\end{equation}

\section{Green-Kubo relation}\label{s:gk}

In this section, we provide  a simple  proof of the Green-Kubo  
relation for the one-dimensional driven lattice gas.  During 
the derivation, we obtain the key identity given in (\ref{ident0}). 
Using this key identity, we can derive  the nonlinear  response 
relation, the fluctuation theorem and the steady state distribution 
as well as the Green-Kubo relation. 

\subsection{Main claim}\label{s:gkmain}

We first define the spatially averaged current per one MCS at time 
$t$, $j(t)$. Because the number of particles  moving from the site   
$x$  to the site $x+1$ from $t$  to $t+1$ is expressed as   
$\sum_{k=0}^{L-1} \Phi_x(\veceta(Lt+k)\to \veceta(Lt+k+1))$, 
$j(t)$ is  written as 
\begin{equation}
j(t)\equiv \frac{1}{L}\sum_{x=0}^{L-1} 
\sum_{k=0}^{L-1}\Phi_x(\veceta(Lt+k)\to \veceta(Lt+k+1)).
\end{equation}
Using the steady state current, $\bar J= \bra j \ketsE$, the 
conductivity, $\sigma$, is defined  as  
\begin{equation}
\sigma \equiv \lim_{E\to 0} \frac{ \bar J}{E}. 
\label{sigdef}
\end{equation}
Furthermore, for the $\tau$-averaged current 
\begin{equation}
J_\tau ([\veceta])\equiv \frac{1}{\tau} \sum_{t=0}^{\tau-1} j(t), 
\label{cdef2}
\end{equation}   
its fluctuation  is characterized by 
\begin{equation}
B \equiv \lim_{\tau\to\infty} \frac{1}{2}\tau L
\bra  J_{\tau}([\veceta])^2  \ketsEz. 
\end{equation}
Then, the Green-Kubo relation connects the conductivity, $\sigma$,   
with the current fluctuation, $B$, by  
\begin{equation}
\sigma= \frac{B}{T}.
\label{gk}
\end{equation}

\vskip3mm
{\it Proof:} 
We prove (\ref{gk}) using a rather special situation. 
Suppose that the statistical distribution of the configuration 
$\veceta$ at $t=0$ is  canonical, $\pc(\veceta)$,  and the  
driving force $E$ is turned on at $t=0$. In this transient process,  
the probability of a history $[\veceta]$  is given by 
\begin{equation}
\PtE([\veceta]) =  \pc(\veceta(0))
T(\veceta(0)\to\veceta(1))\cdots T(\veceta(L\tau-1)\to\veceta(L\tau)) .
\end{equation}
Then, from  (\ref{derh}), (\ref{LDB}) and (\ref{cdef2}), we can 
derive 
\begin{equation}
\frac{\PtE([\veceta])}{\PtE(\tilde{[\veceta]})}
= \e^{\beta E \tau L J_\tau([\veceta])},
\label{defa}
\end{equation}
where $\tilde{[\veceta]}$ represents the time reversed history  
($\veceta(L\tau),\veceta(L\tau-1),\cdots,\veceta(0)$).

Because  the statistical average of a history dependent quantity $A$ 
by the probability $\PtE([\veceta])$ is written as 
\begin{equation}
\bra A \kettE\equiv\sum_{[\veceta]} \PtE ([\veceta]) A([\veceta]),
\end{equation}
the following identity is derived, using (\ref{defa}):
\begin{eqnarray}
\bra A\kettE
  &=&   \sum_{[\veceta]} \e^{\beta E \tau L J_\tau([\veceta])} 
 \PtE (\tilde{[\veceta]}) A([\veceta]),\nonumber \\
 &=&  \sum_{[\veceta]} \e^{-\beta E \tau L J_\tau([\veceta])}
 \PtE  ([ \veceta]) \tilde A([\veceta]), \nonumber \\
 &=& \bra \e^{-\beta E \tau L  J_\tau} \tilde A \kettE,
\label{ident0}
\end{eqnarray}
where $\tilde A([\veceta])\equiv A(\tilde{[\veceta]})$,  
and we have used $J_\tau(\tilde{[\veceta]})=-J_\tau([\veceta])$ 
in deriving the second line.  The expression (\ref{ident0}) 
is the {\it key identity} to derive the universal relations in 
 linear response theory.  We emphasize that  this key identity 
results from the local detailed balance condition (\ref{LDB}) 
and that  this identity is valid even outside the linear response 
regime.

Here, Setting $A=J_\tau$ in (\ref{ident0}), we obtain 
\begin{equation}
\bra J_\tau \kettE =\frac{1}{2}\bra (1-\e^{-\beta E \tau L J_\tau})  
J_\tau \kettE, 
\label{ident}
\end{equation}
which  is called the nonlinear response relation. 
Yamada and Kawasaki initially derived it  \cite{YK}  
and then  Kawasaki and Gunton generalized it  \cite{KG}.  
Finally, by expanding 
the right hand side of (\ref{ident}) with respect to $E$, 
(\ref{ident}) becomes 
\begin{equation}
\bar J= \frac{B}{T}E+O(E^2),
\end{equation}
where  in the large $\tau$ limit,  the statistical average in this 
special situation  $\bra  \  \kettE$,  is equal to the average in 
the steady state $\bra \  \ketsE$. This expression provides 
(\ref{gk}) when the limit $E\to 0$ is used. 

\vskip3mm
{\it Remark:}
One may note that a long time tail of current fluctuation  
yields an anomalous contribution to the Green-Kubo relation.  
This long time tail is mainly due to  momentum conservation 
\cite{Pomeau}. In the one-dimensional driven lattice gas, the 
numerical check indicates that the long time tail of the current 
fluctuation only appear when $E\ne 0$.   (See Refs. \cite{BKS,PS} 
for the case $E\ne 0$.) We conjecture that mode coupling effects 
are vanishing when $E=0$, due to the absence of the momentum 
degrees of freedom. Thus, we cannot discuss a compatibility of 
the long time tail with the Green-Kubo relation within 
the one-dimensional lattice gas. However, it is generally 
understood that a long time tail does not cause the Green-Kubo 
relation to breakdown when we take a long time limit  
with a fixed system size \cite{Murakami}.

\subsection{Steady state distribution}

We can show that the key identity  (\ref{ident0}) yields several 
relations by substituting the appropriate quantities into $A$. 
Through understanding of the fluctuation theorem, originally 
discovered in Ref. \cite{ECM}, the importance of (\ref{ident0}) 
has been recognized \cite{GC,Kurchan,Maes,Crooks}.  Indeed, by 
setting $A([\veceta])=\delta(E\tau L J_{\tau}([\eta])-a)$, and  
by setting $A([\veceta])=1$ in (\ref{ident0}), we obtain the 
fluctuation theorem.  It is noteworthy that the connection of the 
fluctuation theorem with the nonlinear response relation (\ref{ident}), 
which is also derived from  (\ref{ident0}), was noted by Crooks 
\cite{Crooks}.

Furthermore, as  indicated in Ref. \cite{Crooks}, setting 
$A([\veceta])=\delta (\veceta-\veceta(\tau))$ in  the key 
identity  (\ref{ident0}), the steady state distribution function 
is obtained as 
\begin{equation}
\ps_E(\veceta)= \lim_{\tau \to \infty} 
\bra \e^{-\beta E \tau L J_\tau} \delta (\veceta-\veceta(0)) \kettE.
\label{ps1}
\end{equation}
(\ref{ps1}) can be  rewritten 
so that the deviation from the equilibrium distribution function,  
$\pc(\veceta)$,  is explicit. That is, 
\begin{equation}
\ps_E(\veceta)=\pc(\veceta)\lim_{\tau \to \infty} 
\bra \e^{-\beta E \tau L J_\tau}\ket_{E,\veceta(0)=\veceta}, 
\label{ps2}
\end{equation}
where $\bra\ \ket_{E,\veceta(0)=\veceta}$ represents the average 
with the initial condition of $\veceta(0)=\veceta$ when $t=0$.  
Here, (\ref{ps2}) can be also rewritten as 
\begin{equation}
\ps_E(\veceta)=\pc(\veceta)\lim_{\tau \to \infty} 
\e^{\sum_{n=1}^{\infty}\frac{1}{n!}(-\beta E \tau L)^n 
\bra (J_\tau)^n \ket_{E,\veceta(0)=\veceta}^{\rm cum}} , 
\label{ps4}
\end{equation}
where $\bra\ \ket_{E,\veceta(0)=\veceta}^{\rm cum}$ denotes the 
cummulant.   It should be noted that  the expressions of the 
steady state distribution (\ref{ps2}) and (\ref{ps4})  are valid 
even outside  the linear  response regime. 

Particularly, in the linear response regime, (\ref{ps4}) is reduced to 
\begin{equation}
\ps_E(\veceta)= \pc(\veceta)\lim_{\tau \to \infty} 
\e^{-\beta E \tau L \bra J_\tau\ket_{E=0,\veceta(0)=\veceta}+O(E^2)}. 
\label{ps3}
\end{equation}
This expression indicates that the deviation of $\ps_E(\veceta)$ from 
the equilibrium distribution,  $\pc(\veceta)$, is represented by  
an entropy production.  Zubarev and McLennan found a similar 
expression  by studying  the nonequilibrium steady state 
distribution \cite{Zubarev,Maclennan}.  

Furthermore, when the contribution of the order $E^2$ is 
considered, $\ps_E(\veceta)$ can be expressed as 
\begin{equation}
\ps_E(\veceta)=
 \pc(\veceta)\lim_{\tau \to \infty} 
\e^{-\beta E \tau L \bra J_\tau\ket_{E,\veceta(0)=\veceta}
+\frac{1}{2}(\beta E \tau L)^2 \bra (J_\tau)^2\ket_{E=0,\veceta(0)=\veceta}
+O(E^3)}. 
\label{ps5}
\end{equation}
It is interesting to see the fact that the existence of 
the Green-Kubo relation guarantees  the convergence of the 
right-hand side of the expression (\ref{ps5})  with the limit 
$\tau\to \infty$.  

\subsection{Reciprocity relation}

To simply demonstrate the reciprocity relation \cite{ons1,ons2}, 
let us consider a rather artificial model that consists of two 
types of particles labeled,  (X) and (Y).  (X) particles  are 
driven by $E_1$ while (Y) by $E_2$. In this situation,  
$Q(\veceta\to\veceta')$ defined in (\ref{derh}) becomes  
\begin{equation}
Q(\veceta\to\veceta')=H_0(\veceta')-H_0(\veceta)
- E_1 \sum_{x=0}^{L-1} \Phi_x^{\rm X}(\veceta \to \veceta') 
- E_2 \sum_{x=0}^{L-1} \Phi_x^{\rm Y}(\veceta \to \veceta'),
\label{ons1}
\end{equation}
where  $\Phi_x^{\rm X}$ and  $\Phi_x^{\rm Y}$ are the net numbers 
of particles labeled (X) and (Y) that hop from $x$ to $x+1$ for 
the configuration change $\veceta \to \veceta'$, respectively. 
Then, we define the spatially averaged currents per one MCS at time 
$t$,  $j^{\rm X}(t)$ and  $j^{\rm Y}(t)$ as 
\begin{eqnarray}
j^{\rm X}(t)&\equiv&\frac{1}{L}\sum_{x=0}^{L-1} 
\sum_{k=0}^{L-1}\Phi_x^{\rm X}(\veceta(Lt+k)\to \veceta(Lt+k+1)), 
\label{jX}\\ 
j^{\rm Y}(t) &\equiv& \frac{1}{L}\sum_{x=0}^{L-1} 
\sum_{k=0}^{L-1}\Phi_x^{\rm Y}(\veceta(Lt+k)\to \veceta(Lt+k+1)).
\label{jY}
\end{eqnarray}
Using (\ref{jX}) and (\ref{jY}),  the $\tau$-averaged currents 
are defined as  
\begin{eqnarray}
J_\tau^{\rm X}([\veceta]) &\equiv& \frac{1}{\tau} \sum_{t=0}^{\tau-1} 
j^{\rm X}(t),\\ 
J_\tau^{\rm Y}([\veceta])&\equiv& \frac{1}{\tau} \sum_{t=0}^{\tau-1} 
j^{\rm Y}(t).
\end{eqnarray}     

When $E_1\ne 0$ and $E_2=0$, we derive  the key identity in this 
situation:  
\begin{equation}
\bra A\ket^{\rm tr}_{E_1} = \bra \e^{-\beta E_1 \tau L 
J^{\rm X}_\tau} \tilde A \ket^{\rm tr}_{E_1}.
\label{keyons}
\end{equation}
Setting 
$A=J^{\rm Y}_\tau$ and expanding (\ref{keyons}) with respect to $E_1$,  
we obtain 
\begin{equation}
\bra  j^{\rm Y}\kets_{E_1,E_2=0} 
=\frac{1}{2}\beta E_1 \tau L 
\bra  J^{\rm X}_\tau J^{\rm Y}_\tau\kets_{E_1=0}+O(E_1^2).  
\label{ons3}
\end{equation}
Here, we define  the  conductivity, $\sigma_{12}$, which 
characterizes the transportation of type (Y) particles caused by 
the driving force $E_1$ (only applied to type (X) particles), as 
\begin{equation}
\sigma_{12} \equiv \lim_{E_1\to 0} \frac{1}{E_1}
\bra  j^{\rm Y}\kets_{E_1,E_2=0}. 
\end{equation}
Then, (\ref{ons3}) can be rewritten as  
\begin{equation}
\sigma_{12}  
=\frac{1}{2}\beta\tau L 
\bra  J^{\rm X}_\tau J^{\rm Y}_\tau\kets_{E_1=0,E_2=0}.  
\label{rrons}
\end{equation}

Considering the opposite case, when $E_2\ne 0$ and $E_1=0$,  we 
define the conductivity, $\sigma_{21}$, as 
\begin{equation}
\sigma_{21} \equiv \lim_{E_2\to 0} \frac{1}{E_2}
\bra  j^{\rm X}\kets_{E_1=0,E_2}. 
\label{rr1}
\end{equation}
In a similar way as deriving (\ref{rrons}), we can also derive
\begin{equation}
\sigma_{21}  
=\frac{1}{2}\beta\tau L 
\bra  J^{\rm Y}_\tau J^{\rm X}_\tau\kets_{E_1=0,E_2=0}.
\end{equation} 
{}From the trivial identity
\begin{equation}
\bra  J^{\rm X}_\tau J^{\rm Y}_\tau \kets_{E_1=0,E_2=0}=
\bra  J^{\rm Y}_\tau J^{\rm X}_\tau\kets_{E_1=0,E_2=0},
\end{equation}
we obtain the reciprocity relation between the conductivities as  
\begin{equation}
\sigma_{12}=\sigma_{21}. 
\label{ons7}
\end{equation}

\section{Fluctuation-response relation}
\label{s:frr}

The fluctuation-response relation (or the fluctuation-dissipation 
relation)  represents a relation that connects a time dependent 
response to a time dependent external force with a time-correlation 
function in  equilibrium.  There are two types of 
fluctuation-response  relations.  One, which is regarded as an 
extension of the Green-Kubo relation, is when the external force 
is a  time-dependent driving force.  The other is when the external 
force is generated from a time dependent potential and is regarded 
as an extension of the fluctuation relation that can be derived  
within  equilibrium statistical mechanics.

\subsection{Main claim (time-dependent driving force)}

Let us consider the case that  the strength of the driving force 
$E(t)$ changes at every  MSC. When the strength is sufficiently 
weak, the time-dependent averaged current  $\bra j(t) \ket $ (with 
an initial condition given at $t=-\infty$) is expressed by
\begin{equation}
\bra j(t) \ket =\sum_{s=0}^\infty R(s) E(t-s)+O(E^2),
\label{jres4}
\end{equation}
where $R(s)$  is independent of $E$, and  is   called a 
{\it response function}.  Note that, from the causality, $R(t)=0$ 
for $t <0$. In a special situation when  the driving force $E$ is 
turned on at $t=0$; that is, $E(t)=E$(=const.) for $t \ge 0$ and 
$E(t)=0$ for $t < 0$,   the relation (\ref{jres4}) becomes  
\begin{equation}
\bra j(t) \kettE=E  \sum_{s=0}^{t} R(s)+O(E^2).
\label{Rdef}
\end{equation}
On the other hand, setting $A=j(\tau-1)$ in the key identity   
(\ref{ident0}) and using  $\tilde A=-j(0)$,  
we can derive
\begin{equation}
\bra j (\tau-1)\kettE=\beta E  \sum_{s=0}^{\tau-1} \bra j(s) j(0)
\kets_{E=0}-\bra j (0)\kettE+O(E^2). 
\label{Rcal0}
\end{equation} 
In the case  $\tau=1$, this expression yields
\begin{equation}
\bra j (0)\kettE=\frac{1}{2}\beta E   \bra j(0)^2\kets_{E=0}+O(E^2).
\label{j0l}
\end{equation} 
Substituting this into (\ref{Rcal0}), we obtain
\begin{equation}
\bra j (\tau-1)\kettE=\beta E  \sum_{s=0}^{\tau-1}
\theta(s) \bra j(s) j(0)\kets_{E=0}+O(E^2),
\label{Rcal}
\end{equation} 
where $\theta(s)=1$ for $s \ge 1$, and  $\theta(0)=1/2$.
When  we compare (\ref{Rdef}) and (\ref{Rcal}),  using the 
time correlation function
\begin{equation}
C(t)\equiv \bra j(t) j(0)\kets_{E=0},
\end{equation}
we find the relation for $t\ge 0$:  
\begin{equation}
\theta(t) C(t)=T R(t). 
\label{frr:noneq0}
\end{equation}
(\ref{frr:noneq0})  can be  rewritten as
\begin{equation}
C(t)=T (R(t)+R(-t))
\label{frr:noneq}
\end{equation}
for all $t$. 
The relation (\ref{frr:noneq}) is called  the  
{\it fluctuation-response relation},  which is a  dynamic 
extension of the Green-Kubo relation.

\subsection{Main claim (time-dependent potential)}

The relation between the time correlation function and the response 
function can also be derived when a perturbation potential function 
becomes time dependent. Specifically, let us consider a  situation 
when a time-dependent potential
\begin{equation}
V_x \equiv v(t) \sin \frac{2\pi x}{L}
\label{pote}
\end{equation}
is applied to the system. That is, the equilibrium Hamiltonian 
(\ref{eqhami}) is  modified as 
\begin{equation}
H_{V}(\veceta,t)\equiv H_0(\veceta)+ \sum_{x =1}^{L}\eta_x V_x. 
\end{equation}
Here, the amplitude of the density fluctuation with wavenumber  
$2\pi/L$ is 
\begin{equation}
\hat \rho(t) \equiv \sum_{x=1}^{L} \eta_x(Lt)\sin\frac{2\pi x}{L}.  
\label{r1def}  
\end{equation}
Using (\ref{r1def}),  
we define the response function, $R_\rho(t)$, by the relation  
\begin{equation}
\bra \hat \rho(t)\ket=-\sum_{s=0}^\infty R_\rho(s) v(t-s)+O(\Delta^2),
\label{rhores4}
\end{equation}
with the initial condition given at $t=-\infty$.  Note that  
$\bra \hat \rho(t)\ket$ denotes the average in the situation.   
$R_\rho(t)=0$ for $t < 0$ due to the causality. Then, for $t \ge 0$, 
we can  derive the relation 
\begin{equation}
C_\rho(t)=T R_\rho(t), 
\label{frrRC}
\end{equation}
where  
\begin{equation}
C_\rho(t) \equiv \frac{1}{2} \bra (\hat \rho(t)- \hat \rho(0))^2\ketsEz. 
\label{cdef}
\end{equation}  
The relation (\ref{frrRC}) is also called the {\it fluctuation-response 
relation}.

We provide the proof of (\ref{frrRC}) using  a special
situation 
\begin{eqnarray}
v(t) &=& \Delta \qquad {\rm for} \quad t \ge 0, \nonumber \\
          &=& 0 \qquad {\rm for} \quad t < 0 .
\end{eqnarray}
In this case,
the relation (\ref{frrRC}) is explicitly rewritten as
\begin{equation}
\bra \hat \rho(t)- \hat \rho(0) \kettV=-
\frac{\Delta}{2T}
\bra ( \hat \rho (t)-\hat \rho(0) )^2 \ketsEz 
+O(\Delta^{2}), 
\label{frr}
\end{equation}
where $\bra \ \kettV$ denotes the statistical average by   
the canonical
distribution of $v=0$ at $t=0$ and by the transition probability $T$ 
with $v=\Delta$ for $t \ge 0$.  (Note that the canonical distribution 
is realized at $t=0$, because the initial condition is given  at 
$t=-\infty$.) The form (\ref{frr})  can also be derived using a  
method similar to that used for the derivation of the Green-Kubo 
relation.  Indeed, by repeating a  procedure similar to those in 
Subsection \ref{s:gkmain}, we can obtain the following key identity 
for an arbitrary history dependent quantity, $A([\veceta])$:  
\begin{equation}
\bra A \kettV=
\bra \e^{\beta \Delta(\hat\rho(t)-\hat \rho(0))}\tilde{A}\kettV.
\label{ft2}
\end{equation}
Setting $A([\veceta])=\hat \rho(t)-\hat \rho(0)$ and expanding the 
right hand side with respect to $\Delta$, we obtain  (\ref{frr}).  

\noindent
{\it Remark:} 
Although (\ref{frr:noneq}) and (\ref{frrRC}) have similar forms, it  
should be noted that  $R_\rho$ does not represent the current 
response to the time-dependent potential perturbation. When we  
consider the current response to the time-dependent potential  
perturbation, we define the response function of the time difference 
of $\hat \rho$, $R_{\rho}^{\rm d}(t)$, by
\begin{equation}
\bra \hat \rho(t+1)-\hat \rho(t)\ket
=-\sum_{s=0}^\infty R_\rho^{\rm d}(s) \Delta(t-s)+O(\Delta^2).
\end{equation}
Note that 
the time difference of $\hat \rho$ is related to the current due to 
the continuum equation of the density field.  Using the response 
function, $R_{\rho}^{\rm d}(t)$,  the relation  (\ref{frrRC}) can 
be written as 
\begin{equation}
C_\rho(t+1)-C_\rho(t)=T R_\rho^{\rm d}(t).  
\label{frrRC-d}
\end{equation}

\noindent
{\it Remark2:}
In addition to the proof using the key identity (\ref{ft2}), 
which is directly due to  the local detailed balance condition,  
there is another method to prove (\ref{frr}) \cite{Yoshimori}.  
Let $\pc_V(\veceta) $ be the canonical distribution of the system 
with the perturbation $V_x$ of $v(t)=\Delta$. It is easy to find
\begin{equation}
\pc_V(\veceta) =\left(1-\beta\Delta \hat\rho\right)\pc(\veceta) 
+O(\Delta^2).
\label{frr2:pc}
\end{equation}
Using this,  we obtain
\begin{eqnarray}
\bra \hat\rho(t)-\hat\rho(0) \ket^{\rm tr}_{V}
&=&\bra ( \hat\rho(t)-\hat\rho(0) ) 
\left(1+\beta\Delta\hat\rho(0) \right) 
\kets_{E=0,V}+O(\Delta^2) \nonumber \\
&=& \beta\Delta 
\bra\hat\rho(t)\hat\rho(0)\kets_{E,V=0}
-\beta\Delta\bra\hat\rho^2\kets_{E,V=0}+O(\Delta^2).
\label{frr2:siki}
\end{eqnarray}
(\ref{frr2:siki}) 
is equivalent to (\ref{frr}). In this method, we only use  the 
condition that the stationary distribution of the system with the 
potential modulation is written as the canonical distribution.
It means that (\ref{frr}) can be derived without using the local 
detailed balance condition, unlike the case of the Green-Kubo relation.

\subsection{Fluctuation relation}

When the limit $t\to\infty$ is taken, (\ref{frr}) becomes
\begin{equation}
\bra \hat \rho \kets_{E=0,V}=-\frac{\Delta}{T}
\bra  \hat \rho ^2 \kets_{E=0,V=0} +O(\Delta^{2}).
\label{ifrr}
\end{equation}
As shown in the above proof ({\it Remark2}), this kind of relation  
generally holds due to equilibrium statistical mechanics. 
The fluctuation-response relation (\ref{frr}) is a 
dynamic extension of this relation.

Here, we emphasize that (\ref{ifrr}) can be understood from a  
thermodynamic point of view.  When $L$ is sufficiently large, 
$V_x$ of $v(t)=\Delta$ defined by (\ref{pote}), slowly varies  
in $x$. Thus,  $V_x$ and $\bra\eta_x\kets_{E=0,V}$ are regarded as  
smooth functions $V(x)$ and $\rho(x)$. Then, from a thermodynamic 
variational principle and using the free energy density for the 
equilibrium lattice gas, $f(T,\rho)$, the most probable density 
profile $\rho(x)$ under the influence of the slowly varying potential 
$V(x)$ is obtained as the minimizer of the  functional 
\begin{equation}
F(\{\phi(x)\})= \int_0^L dx [f(T,\phi(x))+V(x)\phi(x)], 
\label{fe}
\end{equation}
under the constraint condition 
\begin{equation}
\int_0^L dx \phi(x)= L \bar \rho.
\end{equation}
From this variational principle, $\rho(x)$ is determined so as to  
satisfy 
\begin{equation}
\mu(T,\rho(x))+V(x)-\lambda=0,
\label{ein1}
\end{equation}
where $\lambda$ is the Lagrange multiplier due to the constraint 
condition, and $\mu(T,\rho)$ is the chemical potential defined by
\begin{equation}
\mu\equiv \pder{f(T,\rho)}{\rho}.
\label{mudef}
\end{equation}
Differentiating the relation (\ref{ein1}) with respect to $x$, 
we obtain
\begin{equation} 
\der{\rho(x)}{x}
\left. \pder{\mu(T,\rho)}{\rho}\right\vert_{\rho=\rho(x)} 
 +\der{V(x)}{x}=0.
\label{ein222}
\end{equation}
Substituting (\ref{pote}) into (\ref{ein222}), and using (\ref{r1def}),
we derive 
\begin{equation}
\bra \hat \rho \kets_{E,V}=- \frac{L\Delta }{2}
\left( \pder{\mu(T,\bar \rho)}{\bar \rho} \right)^{-1}
+O(\Delta^2), 
\label{hen}
\end{equation}
where $\bar\rho$ is the averaged density. 

Now, let us recall that the thermodynamic variational principle
is derived from the Einstein-Boltzmann formula for the probability 
of the density profile: 
\begin{equation}
{\rm Prob}(\{\rho(x)\})
\simeq  \delta\left(\int_0^L dx (\rho(x)-\bar\rho)\right)\e^{
-\beta F(\{\rho(x)\}) }. 
\label{b-e2}
\end{equation}
When $V(x)=0$,  setting $\rho(x)=\bar\rho+\delta\rho(x)$ and 
expanding $f(T,\rho(x))$ with respect to $\delta\rho(x)$,  we obtain
\begin{equation}
{\rm Prob}(\{\delta\rho(x)\})
\simeq    \delta\left(\int_0^L dx \delta\rho(x)\right)
\e^{ -\frac{\beta }{2}\left(\frac{\partial \mu(T,\bar\rho)}
{\partial \bar\rho}\right) \int_0^L dx (\delta\rho(x))^2}, 
\label{b-e}
\end{equation}
using the definition of the chemical  potential (\ref{mudef}). 
The expression (\ref{b-e}) leads to the fluctuation relation 
\begin{equation}
\left( \frac{\partial \mu(T,\bar\rho)}{\partial \bar\rho}\right)^{-1}=
\frac{\chi}{T},
\label{sfrr}
\end{equation}
where $\chi $ is the intensity of density fluctuations defined by 
\begin{equation}
\chi\equiv \Delta x \bra (\delta \rho(x))^2 \kets_{E=0}
\label{chidef:naive}
\end{equation}
in the continuum description. Here, $\Delta x$ is the length much 
longer than the microscopic correlation length, but much shorter 
than the system size. $\rho(x)$ represents the coarse-grained 
density over the region of the length $\Delta x$.  

Substituting this result into (\ref{hen}), we obtain
\begin{equation}
\bra \hat \rho \kets_{E=0,V}=- \frac{L\Delta}{2T}\chi 
+O(\Delta^2). 
\label{hen2}
\end{equation}
Thus, (\ref{ifrr}) is  the fluctuation relation, if 
we can prove the relation
\begin{equation}
\chi =  \lim_{L \to \infty} \frac{2}{L}  
\bra \hat \rho^2 \ketsEz. 
\label{chir}
\end{equation}

\vskip3mm
{\it Proof of (\ref{chir}):} 
Let us define the intensity of the density fluctuations $\chi$ 
more explicitly than that in the continuum description given  
in  (\ref{chidef:naive}).  Consider a coarse-grained density of 
the subsystem with the size $\ell$: 
\begin{equation}
\rho_\ell\equiv \frac{1}{\ell} \sum_{x=1}^{\ell} \eta_x.
\end{equation}
Then, using the quantity 
\begin{equation}
d_{\ell,L}\equiv \ell \bra (\rho_\ell-\bar \rho) ^2 \ketsEz, 
\end{equation}
the intensity of density fluctuations,  $\chi$,  should be defined as  
\begin{equation}
\chi\equiv \lim_{\ell \to \infty} \lim_{L/\ell \to \infty} d_{\ell,L}.
\label{dchi}
\end{equation}

Based on this definition, we now derive (\ref{chir}).  Notice the 
asymptotic relation
\begin{equation}
d_{\ell,L}\simeq \chi \frac{L-\ell}{L} 
\label{asymp}
\end{equation}
for $1 \ll \ell \ll  L$.  The derivation of (\ref{asymp}) is 
explained as follows.

Let $N_\ell=\rho_\ell \ell $ and $\bar N_\ell=\bar \rho \ell $. Then, 
according to the central limit  theorem, the  probability of $N_\ell$ 
approaches to the form
\begin{eqnarray}
{\rm Prob}(N_\ell) &\simeq & \e^{
- \frac{(N_\ell-\bar N_\ell)^2}{2 \chi \ell}
-\frac{(N-N_\ell-(N-\bar N_\ell))^2}{2 \chi (L-\ell)}
}  
\nonumber \\
&=&
\e^{
- \frac{(N_\ell-\bar N_\ell)^2}{2 \chi \ell}
-\frac{(N_\ell-\bar N_\ell)^2}{2 \chi (L-\ell)} 
}, 
\end{eqnarray}
when  $\ell$ and $L-\ell$ become large. 
From this, the probability of $\rho_\ell$ is written as 
\begin{equation}
{\rm Prob}(\rho_\ell)\simeq  \e^{ 
- \frac{ \ell L (\rho_\ell-\bar \rho)^2 }{2 \chi (L-\ell)} 
}.
\end{equation}
This implies (\ref{asymp}). 

Using this asymptotic relation, we obtain
\begin{equation}
\chi =  2 \lim_{L \to \infty} d_{L/2,L}. 
\label{chir00}
\end{equation}
Furthermore,  we can derive   
\begin{equation}
\lim_{L \to \infty} d_{L/2,L} = \lim_{L \to \infty} 
{L}  \bra \hat \rho^2 \ketsEz, 
\label{chir01}
\end{equation}
using a straightforward calculation. Combining (\ref{chir00}) 
and (\ref{chir01}), we obtain the expression  (\ref{chir}).

\section{Einstein relation}\label{s:ein}

As is well-known, one of the celebrated papers written by Einstein 
in 1905 is regarded as genesis of  linear response theory 
\cite{Einstein1,Einstein2}.  Even in  many-body systems, if a 
diffusion constant can be  properly defined, it is connected 
with the conductivity in an extended form of the original Einstein 
relation for one-body systems. In this section, we derive the 
Einstein  relation for the one-dimensional driven lattice gas.  
We also explain the interrelations among the Green-Kubo relation, 
the fluctuation-response relations, and the Einstein relation.  
At the end of this section, we discuss the Einstein relation from 
the thermodynamic point of view. 

\subsection{Main claim} 

We begin by defining  the density diffusion constant,  $D$.  
We first assume that the most probable process of the density 
field is described by a diffusion equation.  On this assumption, 
$D$ should be defined as  a coefficient of the diffusion term 
that appears in the equation. As an example, we consider the 
time evolution of the density field after  the slowly varying 
weak potential is turned on at $t=0$. When the most probable 
behavior in this situation can be described by the diffusion 
equation with the diffusion constant $D$, the response function 
$R_\rho(t)$, given in (\ref{rhores4}), can be expressed as
\begin{equation}
R_\rho(t)=R_\rho(\infty)
\left(1- \e^{-  D \left( \frac{2\pi}{L} \right) ^2 t}\right).
\label{diff}
\end{equation} 
In this paper, we assume that (\ref{diff}) determines 
the value of $D$.

Now, the Einstein relation in this model connects the diffusion 
constant, $D$, with the conductivity, $\sigma$, which is defined 
by (\ref{sigdef}), as 
\begin{equation}
D \chi= \sigma T.
\label{ein}
\end{equation}
We  derive (\ref{ein}) using both the Green-Kubo relation 
(\ref{gk}) and  the fluctuation-response relation (\ref{frr}).

\vskip3mm
{\it Proof:} 
From (\ref{frr}) and (\ref{diff}), we obtain   
\begin{equation}
\bra ( \hat \rho (t)-\hat \rho(0) )^2 \ketsEz 
= 2\bra \hat \rho^2 \ketsEz 
\left[
1-\e^{-D  \left(\frac{2 \pi}{L} \right)^2 t}
\right].
\label{diff2}
\end{equation}
Notice  the  continuity equation 
\begin{equation}
\eta_{x}(Lt) -\eta_{x}(0) = -J_{x,t}+J_{x-1,t}, 
\label{contin}
\end{equation}
with 
\begin{equation}
J_{x,t}\equiv\sum_{k=1}^{tL} \Phi_x (\veceta(k-1)\to\veceta(k)), 
\label{Jxdef}
\end{equation} 
where $\Phi_x(\veceta\to\veceta')$ is given in (\ref{phix}).   
Using (\ref{contin}) and (\ref{Jxdef}), we derive
\begin{equation}
\hat \rho(t) -\hat \rho(0)= \left( \frac{2 \pi}{L} \right)
\sum_{x=1}^L J_{x,t} \cos \frac{2 \pi x}{L}+ O\left(\frac{1}{L^2}\right).
\label{rj}
\end{equation}
Next, we define 
\begin{equation}
B_\infty \equiv \lim_{t \to \infty} \lim_{L \to \infty }B_{1,t}(L)
\label{b1def}
\end{equation}
with 
\begin{equation}
B_{1,t}(L)\equiv 
\frac{1}{Lt}
\bra  \left( \sum_{x=0}^{L-1} J_{x,t} \cos \frac{2 \pi x}{L} \right)^2 
\kets_{E=0},
\label{b1t0}
\end{equation}
where the order of the two limits in (\ref{b1def}) must not be 
exchanged. Then, from  (\ref{chir}), (\ref{diff2}) and (\ref{rj}), 
we find
\begin{eqnarray}
B_\infty  &=&    D \lim_{L \to \infty} \frac{2}{L} 
\bra \hat \rho^2 \ketsEz  
\nonumber \\
&=&   D \chi.
\label{fin1}
\end{eqnarray}
Furthermore,  we can derive  $B_\infty =  B$ because the 
$\bra J_{x,t} J_{y,t} \ketsEz$ is a rapidly decaying 
function of $|x-y|$  $(\ll  L/2)$ at a fixed $t$. Thus,  we obtain
\begin{equation}
B= D \chi.
\end{equation}
Combining this relation with the Green-Kubo relation (\ref{gk}),
we  arrive at the Einstein relation (\ref{ein}). 

\vskip3mm
{\it Remark:}
We comment on the diffusion 
of a marked  particle which is characterized by a tracer diffusion 
constant. The tracer  diffusion constant is not equal to the bulk 
diffusion constant, $D$, represented by (\ref{diff}), and it is not
related to the other quantities presented in this paper.

\subsection{Interrelation}

We argue the interrelations among the Green-Kubo relation 
(\ref{gk}),  the fluctuation-response relation (\ref{frr}),  
and the Einstein relation (\ref{ein}).  To simplify the argument, 
we assume that the time correlation function, $C_\rho(t)$, is 
written in an exponential form as a function of time. That is, 
\begin{equation}
\bra ( \hat \rho (t)-\hat \rho(0) )^2 \ketsEz 
= 2\bra \hat \rho^2 \ketsEz  
\left[
1-\e^{-\Dc  \left(\frac{2 \pi}{L} \right)^2 t}
\right].
\label{diffc}
\end{equation}
Based on this assumption, the fluctuation-response relation 
(\ref{frr}) in the simple form is  
\begin{equation}
\Dc=D. 
\label{frr:d}
\end{equation}
Similar to the derivation of (\ref{fin1}),  we can derive
\begin{equation}
B= \Dc \chi
\label{bdc}
\end{equation}
from the continuity equation. Now, when the fluctuation-response  
relation in the form (\ref{frr:d}) is valid, we find  with the aid 
of (\ref{bdc}) that the Green-Kubo relation (\ref{gk})   and the 
Einstein relation (\ref{ein}) are equivalent. Also, when both the 
relations (\ref{gk}) and (\ref{ein}) are valid, we obtain   the 
fluctuation-response relation (\ref{frr:d}) by using (\ref{bdc}). 
In this way,  we have confirmed that any two of the three relations 
lead to the other relation, when $R_\rho(t)$ and $C_\rho(t)$ have the 
exponentially decaying forms.

For the driven lattice gas, using the diffusion constant, 
$D_{\rm c}$, given in (\ref{diffc}),  Katz et al presented the 
rigorous proofs  of the Green-Kubo relation and the Einstein  
relation \cite{KLS} (although in a less systematic way than that 
presented here).  It should be noted that there is the 
fluctuation-response relation (\ref{frr:d}) behind the adoption 
of this definition of the diffusion constant. Related to this 
issue, it is problematic in a system far from equilibrium 
 whether to define the diffusion constant from a response 
function or a time correlation function, because   for the
one-dimensional driven lattice gas, the fluctuation-response 
relation (\ref{frrRC}) is violated in the nonequilibrium 
steady states \cite{HSIV}.

Below, we demonstrate  another important interrelation;  
the Einstein relation is equivalent to the thermodynamic 
fluctuation relation (\ref{sfrr}), when  the relaxation of the 
density field is described more phenomenologically instead of  
using (\ref{diff}).

Let $\bar J(\rho,E)$  be the averaged particle current in a system  
with density $\rho$ and  driving force $E$.  When the slowly varying 
potential $V_x$ of $v(t)=\Delta$, given in (\ref{pote}),  is applied 
to the system of $E=0$, first   the   current   should   become  
$\bar J(\bra \eta_x \kets_{E=0,V},-dV_x/dx)$. Consequently,  
a diffusive current $ D \bra (\eta_{x+1}-\eta_{x}) \kets_{E=0,V}$ 
should appear so that  the  total current is equal to zero. That is, 
\begin{equation}
\bar J(\bra \eta_x \kets_{E=0,V},-\der{V_x}{x})
-D \bra (\eta_{x+1}-\eta_x) \kets_{E=0,V}=0.
\label{balance}
\end{equation}
Precisely speaking,  this relation is assumed to define 
$D$ phenomenologically.  As explained in  Section \ref{s:frr}, 
$\bra \eta_x\kets_{E=0,V}$ and $V_x$ can be regarded as $\rho(x)$ 
and $V(x)$.  Thus, (\ref{balance}) can be rewritten as  
\begin{equation}
\bar J(\rho,-\der{V}{x})-D \der{\rho}{x}=0.
\label{ein2}
\end{equation}
The expansion of (\ref{ein2}) with respect to $dV/dx$ yields
\begin{equation}
-\sigma \der{V}{x}-D \der{\rho}{x}=0.
\label{ein22}
\end{equation}
Note that $|dV/dx|$ is sufficiently small because $V(x)$  
slowly varies in $x$. Substituting the result of the thermodynamic 
variational principle (\ref{ein1}) into (\ref{ein22}), we obtain
\begin{equation}
\sigma \frac{\partial \mu(T,\rho)}{\partial \rho}  -D =0.
\label{ein3}
\end{equation}
Then,  the thermodynamic fluctuation relation (\ref{sfrr}) leads 
to the Einstein relation (\ref{ein}).  

The above argument states  that the non-triviality of the Einstein 
relation depends on the selection of which  $D$ to use among 
(\ref{diff}), (\ref{diffc}) and (\ref{ein2}).  Also, it is amazing 
to see that the thermodynamic variational principle, the thermodynamic 
fluctuation relation, and linear response theory  are mutually 
related.   

Here, let us recall that the fluctuation-response relation (\ref{frrRC}) 
can be derived without the local detailed balance condition, but 
only if a general Markov process compatible with the statistical 
mechanics is used.  Using this fact, and admitting the phenomenological 
derivation of the Einstein relation,  we can derive the Green-Kubo 
relation.   It implies that all the relations in  linear response 
theory can be understood  without the local detailed 
balance condition. Although the argument is correct, it assumes  
that the density field obeys a linear diffusion equation.  We wish 
to  emphasize that the Green-Kubo relation and the 
fluctuation-response relation are derived using the local detailed 
balance condition even when the density field does not obey a simple 
linear equation.

\section{Green-Kubo relation in the boundary-driven 
lattice gas}\label{s:b}

In this section, we study a different type of nonequilibrium 
lattice gases where the particles are not driven by an  
external driving force in the bulk, but are driven 
by the difference in chemical potential  between two particle 
reservoirs in contact with the  boundaries. Due to  the local 
detailed balance condition imposed on the model,  we can derive 
the Green-Kubo relation in a boundary-driven lattice gas.

\subsection{One-dimensional boundary-driven lattice gas}

Consider the system in which the sites $x=1$ and $x=L$ 
are in contact  with the particle reservoirs  that have the 
chemical potentials $\mul$ and $\mur$.  Specifically, we  
set $\mul=\mu$ and $\mur=\mu+\Delta\mu$. When $\Delta\mu \ne 0$, 
the particles flow in one direction on average,  but when  
$\Delta\mu=0$, the grand-canonical distribution is realized. 

We describe the time evolution of the configuration of the system, 
$\veceta$, by iterating the following three Monte Carlo steps:

In the first step, a particle is created and annihilated at the site 
$x=1$. This process is represented by flipping the value of $\eta_1$
with the probability $\cl(\veceta)$. Here, changing from $\eta_1=0$ 
to $\eta_1=1$ corresponds to  creating a particle, while changing  
from $\eta_1=1$ to $\eta_1=0$ corresponds to  annihilating a particle.  
The flipping probability, $\cl(\veceta)$, must satisfy the condition, 
which is a simple extension of  (\ref{ex0}),  
\begin{equation}
\cl(\veceta) = \cl(\veceta^1)
e^{-\beta[H_0(\veceta^1)-H_0(\veceta)-\mul ((\veceta^1)_1-\eta_1)]}, 
\label{c1def} 
\end{equation} 
where $H_0(\veceta)$ represents the interaction  between particles,  
and   its form is given by (\ref{eqhami}).  Recall that $\veceta^1$ 
denotes the configuration of the system after flipping  the value 
of $\eta_1$  (see Section \ref{s:model} for the usage ).  Logically 
speaking,  the condition (\ref{c1def}) is necessary for the 
transition probability to  satisfy the local detailed balance 
condition in nonequilibrium steady states.  We will check this 
condition in the next subsection. 

In the second step, we provide a time evolution rule for the bulk. 
We randomly  select a nearest neighbor pair $x$, $y$,  and  exchange  
the value of $\eta_x$ and $\eta_y$ with the probability 
$\cb(x,y;\veceta)$ to satisfy 
\begin{equation}
\cb(x,y;\veceta)=\cb(x,y;\veceta^{xy}) 
\e^{-\beta[H_0(\veceta^{xy})-H_0(\veceta)]}. 
\label{c3def} 
\end{equation}
We also define $\cb(x,y;\veceta)=0$ for $|x-y|\ne 1$.  Note that 
this  process is the same as that for the lattice gas without the 
particle reservoirs. 

In the third step, the procedure of the first step is performed 
at the site $x=L$. That is, we change the value of $\eta_{L}$ to 
that of $1-\eta_{L}$ with the probability $\crr(\veceta)$ to satisfy 
\begin{equation}
\crr(\veceta) = \crr(\veceta^L)
e^{-\beta[H_0(\veceta^L)-H_0(\veceta)-\mur ((\veceta^L)_L-\eta_L)]}.  
\label{c2def} 
\end{equation} 

Then, the corresponding transition probabilities 
$\Tl(\veceta\to\veceta')$, $\Tb(\veceta\to\veceta')$ and, 
$\Tr(\veceta\to\veceta')$  are defined as  
\begin{eqnarray}
&&\Tl(\veceta \to \veceta') = \cl(\veceta)  
\quad {\rm for} \quad \veceta'=\veceta^1 \ne \veceta,   \\
&&\Tb(\veceta \to \veceta') = \frac{1}{L-1} \cb(x,y;\veceta)  
\quad {\rm for} \quad \veceta'=\veceta^{xy}\ne\veceta,  \\
&&\Tr(\veceta\to\veceta') = \crr(\veceta)  
\quad {\rm for} \quad \veceta'=\veceta^L \ne \veceta . 
\end{eqnarray}
The transition probabilities $\Tl(\veceta \to \veceta)$,  
$\Tb(\veceta \to \veceta)$, and  $\Tr(\veceta \to \veceta)$ are 
determined from the normalization conditions of the probability
\begin{eqnarray}
&&\sum_{\veceta'} \Tl(\veceta \to \veceta') =1, \\
&&\sum_{\veceta'} \Tb(\veceta \to \veceta') = 1, \\
&&\sum_{\veceta'} \Tr(\veceta \to \veceta') = 1, 
\end{eqnarray} 
respectively.  Notice that 
these expressions of the transition probabilities lead to the 
equalities
\begin{eqnarray}
&&\frac{\Tl(\veceta \to \veceta')}{\Tl(\veceta' \to \veceta)}
= e^{-\beta[H_0(\veceta')-H_0(\veceta)-\mul ((\veceta')_1-\eta_1)]},  
\label{tdb1:gc} \\
&&\frac{\Tb(\veceta \to \veceta')}{\Tb(\veceta' \to \veceta)}
= e^{-\beta[H_0(\veceta')-H_0(\veceta)]}  ,
\label{tdb2:gc} \\
&&\frac{\Tr(\veceta \to \veceta')}{\Tr(\veceta' \to \veceta)}
= e^{-\beta[H_0(\veceta')-H_0(\veceta)-\mur ((\veceta')_L-\eta_L)]}  .
\label{tdb3:gc}
\end{eqnarray}

\subsection{Local detailed balance condition} 

Using the above time evolution rule, a history of configurations 
during $\tau$ MCS is denoted as 
$[\veceta]=(\veceta(0), \veceta(1), \cdots, \veceta(3L\tau))$.  
For a segment of the history $[\veceta]_k=(\veceta(3k), \veceta(3k+1), 
\veceta(3k+2),\veceta(3k+3))$, where $k=0,1,2,\cdots ,  L\tau-1$, 
we define the product of the transition probability, $T([\veceta]_k)$,  
as 
\begin{eqnarray}
T([\veceta]_k)&=& \Tl(\veceta(3k) \to \veceta(3k+1)) 
\Tb(\veceta(3k+1) \to \veceta(3k+2))  \\ \nonumber  
&\times& \Tr(\veceta(3k+2) \to \veceta(3k+3)).
\end{eqnarray}
Associated with $T([\veceta]_k)$, we also define 
\begin{eqnarray}
T^*([\veceta]_k)&=& \Tr(\veceta(3k) \to \veceta(3k+1)) 
\Tb(\veceta(3k+1) \to \veceta(3k+2))  \\ \nonumber  
&\times& \Tl(\veceta(3k+2) \to \veceta(3k+3)).
\end{eqnarray}
The transition probability $T^*([\veceta]_k)$ corresponds to the 
time evolution rule obtained by exchanging the first and third 
steps introduced in the last subsection. Furthermore,  
$[\veceta]_k^{*}=(\veceta(3k+3), \veceta(3k+2), \veceta(3k+1),
\veceta(3k))$  represents the time reversed segment of 
$[\veceta]_k$.  Then, from (\ref{tdb1:gc}), (\ref{tdb2:gc}) 
and (\ref{tdb3:gc}),  we obtain the relation
\begin{equation}
\frac{ T([\veceta]_k)}{T^*([\veceta]_k^*)}
=\e^{-\beta[H_0(\veceta(3k+3))-H_0(\veceta(3k))
-\mul \Phil(k)-\mur \Phir(k)]}, 
\label{bldb}
\end{equation}
where 
\begin{equation}
\Phil(k)\equiv \eta_1(3k+1)-\eta_1(3k),
\end{equation}
and 
\begin{equation}
\Phir(k)\equiv\eta_{L}(3k+3)-\eta_{L}(3k+2).
\end{equation}
$\Phil(k)$ and $\Phir(k)$ are  numbers of particles that flow into 
the system from the particle reservoirs during  the time steps from 
$3k$ to $3k+3$ at $x=1$ and at $x=L$, respectively.

Now, the increment of the particle number at the site $x$ from 
$t=3k+3$ to  $t=3k$ is 
\begin{equation}
\delta\eta_x(k)\equiv \eta_x(3k+3)-\eta_x(3k).
\end{equation}
Because the total particle number is conserved, the following 
relation holds: 
\begin{equation}
\sum^{L}_{x=1}\delta\eta_x(k)=\Phil(k)+\Phir(k). 
\label{hozon0}
\end{equation}
Using this relation,  and recalling that $\mul=\mu$ 
and $\mur=\mu+\Delta\mu$, the equality (\ref{bldb}) can be 
rewritten as    
\begin{eqnarray}
&&\frac{T([\veceta]_k)}{T^*([\veceta]_k^{*}) }
=\e^{-\beta[H_0(\veceta(3k+3))-H_0(\veceta(3k)) 
-\mu\sum_{x=1}^{L}\delta\eta_x(k)-\Delta\mu \Phir(k)]} \nonumber \\
&=&\e^{-\beta[\left(H_0(\veceta(3k+3))-\mu\sum_{x=1}^{L}\eta_x(3k+3)\right)
-\left(H_0(\veceta(3k))
-\mu\sum_{x=1}^{L}\eta_x(3k)\right)-\Delta\mu \Phir(k)]}.  
\label{bldb2}
\end{eqnarray}
This is regarded as the local detailed balance condition in the 
boundary-driven lattice gas.

In the equilibrium case $\Delta\mu=0$, (\ref{bldb2}) leads to 
\begin{equation}
\pgc(\veceta(3k))T([\veceta]_k)=\pgc(\veceta(3k+3))
T^*([\veceta]_k^{*}),
\label{db:gc}
\end{equation}
with  the grand-canonical distribution $\pgc(\veceta)$ 
\begin{equation}
\pgc(\veceta)\equiv \frac{1}{Z}
\label{gc}
\e^{-\beta [ H_0(\veceta)-\mu\sum_{x=1}^{L}\eta_x]},
\end{equation}
where $Z$ denotes the normalization factor. The equality (\ref{db:gc})
corresponds to the detailed balance condition with respect to the 
grand-canonical distribution. This ensures that (\ref{gc}) is the
steady state distribution of the stochastic model.  

\subsection{Main claim}

In order to derive the Green-Kubo relation in this model, we 
assume that the grand-canonical distribution $\pgc(\veceta)$ of the 
chemical potential $\mu$ is realized at $t=0$.  Then, we  consider 
the situation where $\Delta\mu\ne 0$ when $t\ge 0$. In this case, 
the probability of a history $[\veceta]=(\veceta(0), \veceta(1), 
\cdots, \veceta(3 L\tau))$  is given by 
\begin{equation}
\Pgc([\veceta])\equiv \pgc(\veceta(0)) T([\veceta]_{0}) 
\cdots T([\veceta]_{ L\tau-1}).
\end{equation}
The path probability associated with $T^*$ is also defined as
\begin{equation}
\Pgct([\veceta])\equiv \pgc(\veceta(0)) T^*([\veceta]_{0}) 
\cdots T^*([\veceta]_{ L\tau-1}).
\end{equation}
Noting the equality
\begin{equation}
\Pgct(\tilde{[\veceta]})=\pgc(\veceta(L\tau))  
T^*([\veceta]_{ L\tau-1}^*) \cdots T^*([\veceta]_{0}^*),
\end{equation}
and using the local detail balance condition (\ref{bldb2}), we obtain
\begin{equation}
\frac{\Pgc([\veceta])}{\Pgct(\tilde{[\veceta]})} 
= e^{\beta \Delta\mu \tau J_\tau([\veceta])},
\label{defa2}
\end{equation}
with 
\begin{equation}
J_\tau([\veceta])\equiv
\frac{1}{\tau}\sum_{k=0}^{ L\tau-1}\Phir(k).  
\label{tcur}
\end{equation}

Then,  we write the statistical average  in the transient process 
by the probability, $\Pgc([\veceta])$, as 
\begin{equation}
\bra A \ket^{\rm tr}_{\Delta\mu} = \sum_{[\veceta]} \Pgc([\veceta]) 
A([\veceta]),
\label{siki100}
\end{equation}
for a history dependent quantity $A([\veceta])$.  $\bra A 
\ket^{{\rm tr}*}_{\Delta\mu}$  is also defined in the same way, 
replacing $\Pgc([\veceta])$ by  $\Pgct([\veceta])$ in (\ref{siki100}).   

Noting the equality $J_\tau (\tilde{[\veceta]})=-J_\tau([\veceta])$, 
we obtain the key identity
\begin{eqnarray}
\bra A \ket^{\rm tr}_{\Delta\mu} 
&=&
\sum_{[\veceta]} \e^{\beta \Delta\mu \tau J_\tau([\veceta])}
\Pgct (\tilde{[\veceta]}) A([\veceta]), \nonumber \\ 
&=&  
\sum_{[\veceta]} \e^{-\beta \Delta\mu \tau  J_\tau(\tilde{[\veceta]})}
\Pgct (\tilde{[\veceta]}) \tilde A(\tilde{[\veceta]}), \nonumber \\ 
 &=& \bra \e^{-\beta \Delta\mu \tau J_\tau([\veceta])} 
\tilde A \ket^{{\rm tr}*}_{\Delta\mu},
\label{ident02}
\end{eqnarray}
where we have defined $\tilde A([\veceta])=A(\tilde{[\veceta]})$. 

Here, setting $A=J_\tau$ in (\ref{ident02}), taking the limit 
$\tau\to\infty$ and expanding the right hand side of (\ref{ident02}) 
with respect to $\Delta\mu$, we can derive 
\begin{equation}
\bra J_\tau \kets_{\Delta\mu}  = \frac{B}{T}\Delta\mu+O(\Delta\mu^2),
\label{bgk}
\end{equation}
with 
\begin{equation}
B\equiv \lim_{\tau\to\infty}\frac{\tau}{2}
\bra (J_\tau)^2\kets_{\Delta\mu=0}.  
\end{equation}
Note that 
$\bra  \  \ket^{{\rm tr}}_{\Delta\mu}$ and  $\bra  \  \ket^{{\rm tr}*}
_{\Delta\mu}$  are  equal to the average in the steady state 
$\bra \  \kets_{\Delta\mu}$ with  the large $\tau$ limit and a fixed 
system size $L$.  The expression  (\ref{bgk}) provides the 
Green-Kubo relation in  the boundary-driven lattice gas, because  
the current correlation $B$ in the system  under the equilibrium 
condition  is related to the conductivity $\sigma$ in the 
nonequilibrium   in the form $B=\sigma T$, where $\sigma$ is defined 
as 
\begin{equation}
\sigma\equiv \lim_{\Delta\mu\to 0}
\frac{\bra J_\tau \kets_{\Delta\mu}}{\Delta\mu}.
\end{equation}

We remark that the other relations derived in Sections \ref{s:frr} 
and \ref{s:ein}  can be obtained, using the local detailed balance 
condition (\ref{bldb2}).   As far as we know, the Green-Kubo 
relation in this model  was  presented  in a different manner 
in Ref. \cite{Spohn}.  Our derivation method is more pedagogical 
in the sense that the role of the local detailed balance condition 
is  explicit.

\section{Discussion}\label{s:dis}

We have derived the universal relations of linear response 
theory for  bulk-driven and  boundary-driven lattice gases 
in a simple  manner. We have also elucidated the interrelations 
among these universal relations, statistical mechanics and 
thermodynamics. However, in the arguments presented above, 
some important topics  related to  linear response theory 
were not discussed. Among them, there are two that  are 
particularly worth consideration,  the derivation of the 
universal relations from classical or quantum systems, and 
the extension of the universal relations to forms  valid in 
nonequilibrium steady states far from equilibrium. In this 
final section, we present remarks on these two topics. 


\subsection{Microscopic understanding}

It is well known that the universal relations of  linear response 
theory can be derived from a microscopic description in the 
following way.   
(i) The system under 
consideration is assumed to be in equilibrium at $t =-\infty$; 
that is, the distribution of microscopic variables is assumed to be 
canonical at $t=-\infty$. 
(ii) An external force is applied to the system  under the 
assumption that the time evolution obeys the Liouville and von 
Neumann equations.  This assumption implies that the system 
does not interact with other dynamical degrees of freedom. 
(iii) The statistical average of the quantity of interest 
at time $t$ is calculated
perturbativelly up to linear order in the applied external force. 

Within such a framework, it is straightforward to derive the 
linear response relations, but it is difficult to understand the 
physical picture.  For example, let us consider an electric 
conduction system.  In such a system,  energy is continuously  
introduced into  the system when  an electric field is applied. 
Thus, obviously, a steady state is never realized unless the 
system contact with a heat bath. Then, recalling the step (ii) 
described above in the previous paragraph,  it can be claimed that 
the formal calculation is not performed based on the realization
of nonequilibrium steady states, even if the desired formulas can be 
derived by this method.

Furthermore, it should be noted that the time correlation 
(e.g. current correlations)  in the formal perturbation scheme is 
calculated by use of mechanical time evolution equations  without 
identifying the degrees of freedom that constitute the heat bath. 
If the energy dissipation into the heat bath is taken into account 
when the time correlation is calculated, it is plausible that the 
result of the calculation would depend on the nature of the heat bath.    
Indeed, in the stochastic models studied in this paper, the functional 
form of the time correlation depends on the choice of the stochastic 
rule (e.g. (\ref{mt}), (\ref{hb}), and (\ref{sym})), which represent  
the nature of the heat bath,  though the validity 
of the universal relations is independent of this choice.

Considering these points, we find that we do not obtain  a 
clear understanding of nonequilibrium systems, even in the linear 
response regime, when we study systems on the basis of a microscopic 
description.  Here, let us recall our method  of derivation for 
the universal relations  in the stochastic models.  In our 
derivations, the steady state is prepared from the outset, and  
all the relations can be obtained by use of the local detailed 
balance condition. Thus, focusing on the realization of 
nonequilibrium steady states and the local detailed balance 
condition, we suggest that the two problems described below should 
be studied in microscopic systems.  

The first problem is to clarify the conditions under which 
a subsystem exhibiting a nonequilibrium steady state 
is determined in a purely mechanical system despite a continuous 
injection of energy. (Note that a nonequilibrium system with a 
Gaussian thermostat \cite{thermostat} 
is not regarded as a purely mechanical system of the type we wish 
to study.)  Then, we characterize the heat bath by studying the 
mechanism of the energy dissipation, but it is not certain that a 
simple characterization is possible. If we succeed in the proper 
characterization,  the second  problem  is to demonstrate the 
validity of the local detailed balance condition for these states.  
Then, given the  local detailed balance condition, it is 
straightforward to derive universal relations. In this way, 
through the present study, we have identified  a new direction 
in the investigation of  the nature of nonequilibrium systems 
on the basis of a  microscopic description.

\subsection{Extension of the linear response relations}

One may wonder whether it is possible to extend  linear response 
theory  to systems far from equilibrium. A partial  answer to 
this question is provided by the nonlinear  response relation 
given in  (\ref{ident0}),  which provides a starting point to 
connect transport  properties  of systems far from equilibrium  
with  dynamical fluctuations. Indeed, when Kawasaki and Gunton 
derived an expression for the  nonlinear shear viscosity from 
the Liouville equation, the nonlinear response formula \cite{KG},  
which takes essentially  the same form as  (\ref{ident0}), 
played  a key role.  

There is another  approach to the extension of the universal 
relations that does not involve nonlinear response theory.  
In this approach, instead of studying  nonlinear transport  
properties,  the linear response properties  near  nonequilibrium 
steady states are studied by applying small perturbations  as  
probes.  It is believed that there is a close relationship between 
the  linear response properties and the dynamical fluctuation 
properties,  although,  obviously, the universal relations that 
hold within the linear response regime cannot be valid in general 
outside this  regime.  Unlike the nonlinear  response theory 
represented by (\ref{ident0}), this type of extension of the 
universal relations beyond the linear response regime has not 
yet been established. 

As a preliminary step in the attempt to employ this second type of 
approach to construct an extension of the universal relations 
to systems far from equilibrium (i.e. outside the linear response 
regime),  recently, numerical experiments on a two-dimensional 
driven lattice gas were performed \cite{KH}. According to the 
results of  this study, with respect to  the properties of the 
system along  the direction transverse to the external driving force, 
the  linear responses of the system to  perturbation forces  
are related to the  fluctuations in the same form as the linear 
response relations.   Here, it is important to note that in the 
direction transverse to the external driving force, the thermodynamic 
free energy, from  which we can obtain the probability distribution 
for density fluctuations, was constructed using the Maxwell relation 
\cite{HSI,sasatasaki}.  (The Maxwell relation  is  the integrability 
condition yielding the free energy.)  We conjecture that the validity 
of the extended forms of the linear  response relations \cite{KH} is 
related to the existence of the thermodynamic function constructed 
in Ref. \cite{HSI}, recalling the argument given in Section \ref{s:ein}.

Although these numerical results for  states far from equilibrium 
are interesting, our understanding of the extended relations remains 
poor. In particular, there are the following two serious problems. 
First, it is well known that long-range spatial correlations of 
density fluctuations exist in nonequilibrium systems of two or 
more dimensional \cite{Dorfman,Tasaki,Sasa}. These long-range 
correlations are inconsistent with the extensive nature of 
thermodynamic fluctuation theory.  In fact, we do not understand 
how the numerical results of Ref. \cite{KH,HSI} can be reconciled 
with the existence of long-range correlations. (See Ref. 
\cite{sasatasaki} for a related discussion). Second, while  we 
were able to  numerically construct an  extended free energy and 
confirm the validity of the extended linear response relations 
along  the direction transverse to the external driving force, 
it seems more difficult to find  similar relations along  the 
direction parallel to the driving force.   Indeed, the study of a 
one-dimensional driven lattice gas far from equilibrium has revealed 
that the relationship between the fluctuations and the response to 
a probe force  is complicated, because the fluctuations take the 
influence of  hydrodynamic effects \cite{HSIV}.

In order to construct a self-contained  theory of nonequilibrium 
steady states,  we need a unified treatment of thermodynamic and 
hydrodynamic fluctuations.  For example,  thinking optimistically, 
despite the existence of the long-range spatial correlations in 
nonequilibrium steady states,  it might be possible to extract the 
thermodynamic component of the fluctuations by separating out of 
the long-range component.  At present, however, there is no theory 
providing such a treatment.  However, we should mention that the 
additivity principle proposed in Refs. \cite{DLS,DLS2}  might 
provide a useful tool to study this problem, because this principle 
provides an elegant method by  which the thermodynamic part of 
density fluctuations  can be distinguished from  the hydrodynamic 
part. Although the validity of this principle has been confirmed 
only for certain exactly solvable  nonequilibrium models, it would 
be interesting to consider its application to  a wider  class of 
nonequilibrium models and to study how thermodynamic fluctuations 
can be extracted by use of  this additivity principle. 

If we could construct a unified treatment of fluctuations for  
nonequilibrium steady sates far from equilibrium, our next step 
would be to seek  new universal relations between fluctuations and  
response  properties to probe forces in such systems.  In that 
pursuit, and in the construction of a general theory of 
nonequilibrium steady states, the present paper, in which we have 
elucidated the interrelations among the linear response relations, 
statistical mechanics and thermodynamics,  should serve as a useful 
guide.

\section{Acknowledgment}\label{s:ac}

The authors acknowledge T. Harada, H. Tasaki and A. Yoshimori 
for fruitful discussions on linear response theory and 
nonequilibrium steady states.   We also acknowledge G. Paquette 
for helpful comments on the paper. This work was supported by a 
grant from the Ministry of Education, Science, Sports and Culture 
of Japan (No. 16540337) and a grant from Research Fellowships of 
the Japan Society for the Promotion of Science for Young 
Scientists (No. 1711222).



\end{document}